\documentclass[a4paper]{article}
\usepackage{indentfirst}
\usepackage{multirow}
\usepackage{graphicx}
\usepackage{epsfig}
\usepackage{subfigure}
\usepackage{cite}
\usepackage{color}
\usepackage[top=1in, bottom=1in, left=1.25in, right=1.25in]{geometry}

\def\bea{\arraycolsep .1em \begin{eqnarray}}
\def\eea{\end{eqnarray}}

\def\s0#1#2{\mbox{\small{$ \frac{#1}{#2} $}}}
\def\0#1#2{\frac{#1}{#2}}

\usepackage{lmodern}
\usepackage{amsmath, amssymb}
\begin{document}

\setcounter{topnumber}{10}
\setcounter{totalnumber}{50}
\title{Investigation of NWC-induced electron precipitation and theoretical simulation}
\maketitle

\begin{center}
{\sc Z.~X.~Zhang,$^{1)\,,}$
\footnote{e-mail address:
zxzhang@neis.cn
 }\,\,\,
 X.~Q.~Li,$^{2)\,,}$\,\,\,
C.~Y.~Wang,$^{3)\,,}$\,\,\,
L.~J.~Chen$^{4),,}$}
\\
\vspace{0.5cm}

\noindent{\small{1) \it National Earthquake Infrastructure Service, Beijing, China}}\\
\noindent{\small{2) \it Institute of High Energy Physics, Chinese
Academic School, Beijing, China}}\\
\noindent{\small{3) \it Peking University, Beijing, China}}\\
\noindent{\small{4) \it   Department of Physics, University of Texas at Dallas, Richardson, Texas, USA}}
\end{center}
\begin{abstract}
Enhancement of the electron fluxes in the inner radiation belt, which is
induced by the powerful North West Cape (NWC) very-low-frequency (VLF) transmitter, have been
 observed and analyzed by several research groups. However, all of the previous publications
  have focused on NWC-induced $>$100-keV electrons only,
  based on observations from the Detection of
Electro-Magnetic Emissions Transmitted from Earthquake Regions (DEMETER) and the Geostationary
 Operational Environmental Satellite (GOES) satellites.
Here, we present flux enhancements with $30$--$100$-keV electrons related to NWC  transmitter for the first time,
 which were observed by the GOES satellite at night.
  Similar to the $100$--$300$-keV precipitated-electron behavior, the low energy $30$--$100$-keV
  electron precipitation is primarily located east of the transmitter. However,
   the latter does not drift eastward to the same extent as the former,
   possibly because of the lower electron velocity. The $30$--$100$-keV electrons
    are distributed in the $L=1.8$--$2.1$ L-shell range, in contrast to the $100$--$300$-keV electrons which are at $L=1.67$--$1.9$.
This is consistent with the perspective that the energy of the VLF-wave-induced electron flux enhancement decreases
with higher L-shell values.
We expand upon the rationality of the simultaneous enhancement of the $30$--$100$- and $100$--$300$-keV electron
 fluxes through
comparison with the cyclotron resonance theory for the quasi-linear wave-particle
interaction. In addition, we interpret the asymmetry characteristics of NWC electric power distribution in north and south hemisphere by ray tracing model. Finally, we present considerable discussion and show that good agreement exists
 between the observation of satellites and theory.
\end{abstract}

\section{Introduction}\label{intro}
There are many factors that cause high-energy particle acceleration, precipitation, and short-term,
 sharp increases in particle count rates in radiation belts, including the influence of
  ground-based VLF electromagentic (EM) transmitters, lightning and thunderstorms,
   and ground nuclear testing. The primary mechanism of ground-based VLF EM transmission is
    that the emitted EM wave is transmitted across the atmosphere, expands into the ionosphere,
    and interacts with the energetic particles in the radiation belt.
    This interaction accelerates the particles by changing their momentum or scattering their pitch angle,
     which causes them to enter the bounce or drift loss cones $LC_{drift}$.
      As a result, a mass of energetic particles accumulates within a certain L-shell and
       yields particle-flux enhancement, which can be observed by the onboard particle detectors of
        satellites\cite{Reeves1998,Li1997,Su2014,Horne2005,Su2011,Thorne2010}.

Numerous experimental observations and theoretical interpretations of the electron-flux enhancement
 induced in the inner radiation belt by VLF ground-based transmitters have been
 reported.

For example, Kimura {\it et al.} have found strong correlations
between the $0.3$--$6.9$-keV electron fluxes observed by the
EXOS-B satellite and the $0.3$--$9$-kHz VLF wave emitted by the
ground-based transmitter at SIPLE\cite{Kimura1983}. The observed
instantaneous correlation between the VLF signal and the electron
fluxes has been studied\cite{Imhof1983} and interpreted using
wave-particle interaction theory\cite{Inan1985}, which was applied
to a test-particle model of a gyroresonant wave-particle
interaction in order to calculate the precipitation
characteristics of the particle flux induced by a VLF transmitter.
Hence, it was found that the particle precipitation is controlled
by the $\alpha$ distribution near the edge of the loss cone.

The Lualualei (NPM) VLF transmitter has been studied by Inan {\it
et al.}\cite{Inan2007}, who have suggested that the energetic
electrons induced by this transmitter, which are scattered at the
NPM longitude, continue to precipitate into the atmosphere as they
drift toward the South Atlantic Anomaly. Further, Graf {\it et
al.} have compared the precipitating fluxes with predictions based
on ray-tracing analysis of the wave propagation and test-particle
modeling of the wave-particle interaction\cite{Graf2009}. Their
results indicate that the precipitated flux of the $>$100-keV
electrons induced by the NPM transmitter peak at $L\simeq 1.9$.
They have also indicated that the detection rate is related to the
orientation of the Detection of Electro-Magnetic Emissions
Transmitted from Earthquake Regions
(DEMETER)\cite{Parrot2006,Sauvaud2006} particle detector,
obtaining agreement between their observations and theory.

Several studies of the electron-precipitation belts induced by the
North West Cape (NWC) VLF ground-based transmitter have also been
conducted, and the events related to this transmitter are the
focus of this paper. Based on readings from the DEMETER satellite,
Sauvaud {\it et al.} have observed NWC-transmitter-induced
enhancements in the ~100--600-keV $LC_{drift}$ electron fluxes at
$L$ values of 1.4--1.7\cite{Sauvaud2008}. These researchers have
calculated the variation in the energy of the enhanced electron
fluxes in response to changes in $L$ using the first-order
cyclotron resonance theory of wave-particle interaction; hence,
they have obtained results consistent with observation.Further, Li
{\it et al.} have analyzed the energy spectra of the NWC electron
belts precisely using an on-off method, and explained the $\alpha$
range by considering the quasi-linear diffusion equation of
wave-particle theory\cite{Li2012}. Independently of the DEMETER
detection of the NWC-transmitter-induced electron enhancement of
$LC_{drift}$, Gamble {\it et al.} have also examined the data for
$>$100-keV quasi-trapped electron fluxes and reported similar
enhanced electron counts from the $90^{\circ}$ electron telescopes
on the National Oceanic and Atmospheric Administration (NOAA) 15,
16, 17, and 18 Polar-orbiting Operational Environmental Satellite
(POES), for the time period in which the NWC transmitter was
broadcasting\cite{Gamble2008}.


In the early work by Kennel {\it et al.} on the pitch angle scattering of radiation-belt particles,
 the gyroresonant wave-particle interactions were thought to play a crucial role in the
  magnetosphere physics\cite{Kernel1966}. Further, scattering by chorus waves was studied as the dominant cause
   of diffuse auroral precipitation\cite{Thorne2010}. One of the theoretical models for numerically
    calculating the precession of waves inducing energetic particle precipitation is
     the quasi-linear diffusion equation. In particular, Summers {\it et al.} have developed
      a precise formula of quasi-linear diffusion coefficient corresponding to R- or L-mode EM waves
      which have a Gaussian spectral density and propagate in a hydrogen plasma\cite{Summers2005}.
       These researchers have expanded the collisionless Vlasov equation for the particle-flux
        differential function to the second-order perturbation and obtained the diffusion equation form.
         In the inner radiation belt, the particle velocities are generally significantly larger than
          the typical phase velocities of the waves; therefore, pitch angle diffusion plays a dominant
           role in the wave-particle interaction.

In addition, many electron-belt effects induced by man-made VLF
ground-based transmitters can be described using the theoretical
model of wave-particle interaction. For example, Horne {\it et
al.} have studied the mechanism behind the wave-induced electron
acceleration in the outer radiation belt\cite{Horne2005}. Those
researchers have shown that the electrons can be accelerated by EM
waves at frequencies of a few kHz, which can also increase the
electron flux by more than three orders of magnitude over an
observation timescale of 1--2 d.

However, all the previous analyses of the NWC-transmitter-induced
particle precipitation have focused on particles with energies
above 100 keV only. In this paper, we present a more comprehensive
survey of energetic particles for a wider energy spectrum,
including particles with energies below 100 keV. And then we
simulate the wave-particle interaction process for NWC-induced
electron precipitation using quasi-linear diffusion theory, which
considers the $\alpha$ scattering and energy diffusion for
different $L$ values. We compare the simulation
results to the observations from the DEMETER and NOAA satellites.
Finally, we discuss the
limitations of the theory.

\section{Observation}
The NWC ground station is located in the northwest corner of Australia with geographical coordinates of $(21.82^{\circ}S$, $114.15^{\circ}E)$, and geomagnetic coordinates of $(-31.96^{\circ}$, $186.4^{\circ})$, which emits electromagnetic wave with frequency 19.8 kHz, a very narrow bandwidth and a large emission power of 1 MW\cite{Li2012}.

The DEMETER satellite was launched in June 2004 and is a low-altitude
    satellite with onboard detectors to measure local electric and magnetic
     fields and energetic particle populations\cite{Parrot2006,Sauvaud2006}. This satellite,
     having a quasi-Sun-synchronous orbit, travels downward (from north to south)
     during local daytime and upward (from south to north) during local nighttime.
      The orbital period is 102.86 min.

Data on the NWC-induced electron belts have been obtained using
DEMETER and analyzed\cite{Li2012}. These data indicate the
presence of obvious wisp structures, which correspond to a
spectrogram of NWC electrons distributed in five different
$L$-value regions (see Fig. 10 in Ref. [9] for details). Thus, it
has been confirmed that the electron precipitation belts are
caused by EM waves emitted from the NWC transmitter. The $L$
values for the NWC electron belts range from 1.5 to 2.2. The local pitch angle
$\alpha$ for the NWC electron belts measured by the DEMETER
detector is approximately $60$--$110^{\circ}$, which corresponds
to an equatorial pitch angle $\alpha_{eq}$ of $23.5$--$25^{\circ}$
at a latitude of $37^{\circ}$ in the southern hemisphere.

The NOAA satellites are positioned in a polar orbit (inclination
angle: $~99^{\circ}$) at altitudes of 807--854 km. The constituent
particle detectors (Space Environment Monitor; SEM-2), which
monitor the proton and electron fluxes at the satellite altitudes,
consist of total energy detectors (TEDs) and medium energy proton
and electron detector (MEPEDs). An MEPED is composed of eight
solid-state detectors.

\begin{figure}
\centering
\includegraphics[width=38pc]{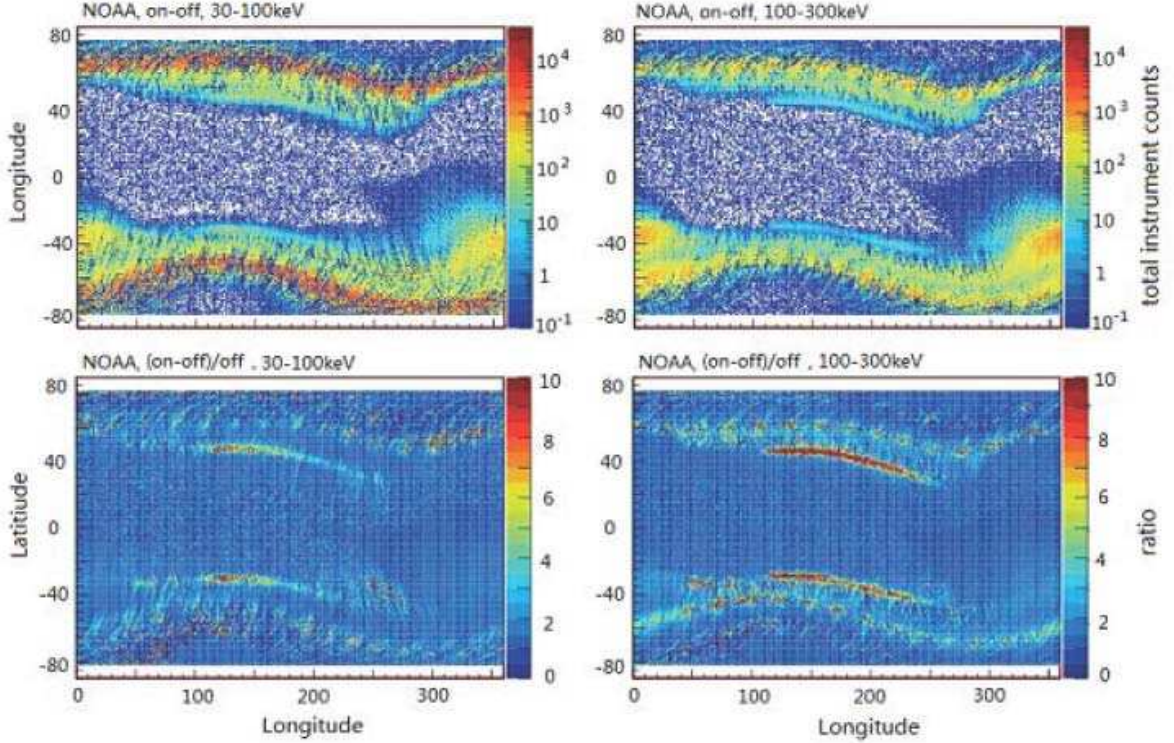}
  \caption{\small Effects of NWC transmission for
  $30$--$100$ and $100$--$300$-keV electron observation data from NOAA satellites.
  (top) Electron counts for 1 January to 31 May 2007 with those for 1 August to 31 December 2007 subtracted.
   The data in each period are the sum of all $30$--$100$- or $100$--$300$-keV electron counts
   from the medium energy proton and electron detectors (MEPEDs) in the $90^{\circ}$ telescopes
    for the given periods. (bottom) Ratio of subtracted data shown above to background electron counts
     for 1 August to 31 December 2007. The NWC was operating normally from 1 January to 31 May 2007. }
  \label{1}
 \end{figure}

In addition to the wisp structure of the NWC-induced electron
precipitation observed by the DEMETER satellite, we have
investigated the observed data for the inner-radiation-belt
electrons using data recorded in 2007 by NOAA satellites.
Following the approach described in Ref.\cite{Gamble2008}, we also
present NWC-induced electron belts obtained from NOAA data in
Fig.\ref{1}. This image shows the significant distributions of the
$30$--$100$- and $100$--$300$-keV electron counts for the
geographic position of the subsatellite point, for signals
obtained in the operating period of 1 January to 31 May 2007. The
background obtained during the non-operational period of 1 August
to 31 December 2007 has been subtracted from these signal data. The
ratios of the subtraction and background electron counts are also
shown in the figure.

Note that waves from the NWC propagate through the ionosphere
primarily at night\cite{Starks2008}. Thus, we combined all the
16-s electron integral flux observations provided by the NOAA
National Geophysical Data Center by summing the observation data
obtained by the NOAA 15,16, 17, and 18 POES at night
(2200--0600UT).

\begin{figure}
\centering
\includegraphics[width=38pc]{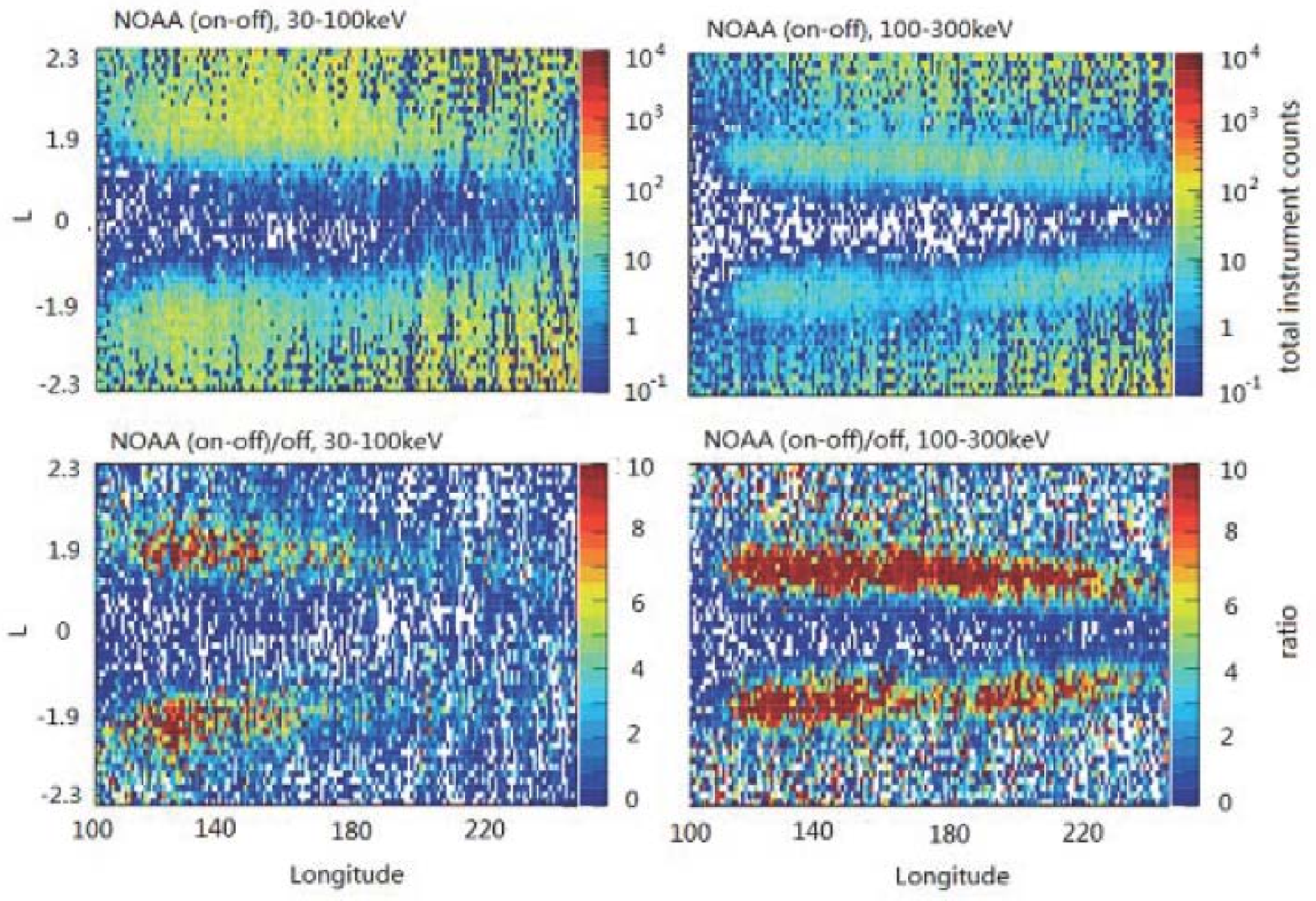}
  \caption{\small Effect of NWC transmission projected in L-shell
  as indicated by $30$--$100$- and $100$--$300$-keV electron observations from NOAA satellites.
   (top) Electrons for 1 January to 31 May 2007, with those for 1 August to 31 December 2007 subtracted.
    The data in each period are the sum of all $30$--$100$- or $100$--$300$-keV electron counts
     from the $90^{\circ}$ MEPEDs for the given periods. (bottom) Ratio of subtracted data shown
      above to background for 1 August to 31 December 2007. The NWC was operating normally
       from 1 January to 31 May 2007. }
  \label{2}
 \end{figure}

Different to Ref.\cite{Gamble2008}, we exhibit the NWC electron
belts in the lower-energy region ($30$--$100$ keV) in the left two
plots of Fig.\ref{1}. These images indicate that the NWC transmitter
induces not only electrons of $>100$ keV, but also those with
energy $<100$ keV. Note that this is the first time that
NWC-induced electron precipitation belts in the lower-energy range
of $30$--$100$ keV have been reported. In Fig.\ref{2}, we also
present the effect of NWC transmission projected into the L-shell
for electron observations in the two energy ranges of $30$--$100$
and $100$--$300$ keV, based on data from the NOAA satellites. The
L-shell values are directly linked to the dynamic mechanism of the
wave-particle interaction, and the L-shell distribution of the NWC
electrons demonstrates that the $30$--$100$-keV electrons are most
likely induced by the NWC transmission in the $1.8$--$2.0$
L-shell, whereas the $100$-$300$-keV electrons are at
$1.7$--$1.9$. This result indicates that lower-energy electrons
can be influenced by NWC EM waves in higher L-shells in the
ionosphere. In addition, Fig.\ref{2} also indicates that the the NWC
electrons drift eastward above the Earth's surface, in accordance
with the basic three-motion theory for charged particles in the
ionosphere, where those motions are gyration, bounce, and drift.
Furthermore, the $30$--$100$- and $100$--$300$-keV NWC electrons
drift from $110^{\circ}$--$160^{\circ}$ and
$110^{\circ}$--$240^{\circ}$ longitude, respectively. This
behavior is likely caused by the different electron velocities.
Note that, the higher the electron velocity, the farther eastward
in the ionosphere the electrons can travel.

\begin{figure}
\centering
\includegraphics[width=38pc]{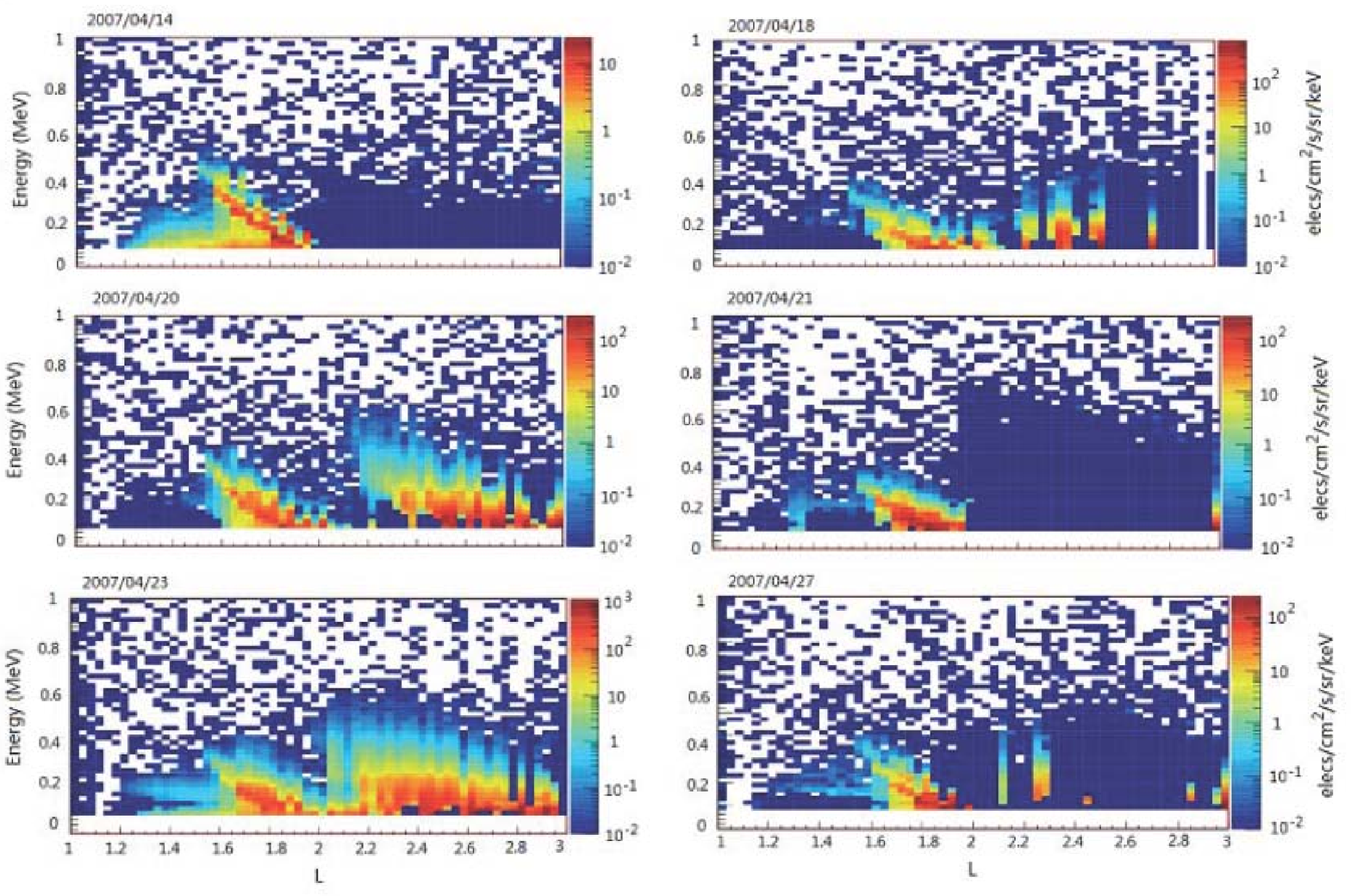}
  \caption{\small Examples of NWC electron precipitation wisp structures
  after background effect is extracted from signal data obtained from IDP on board the DEMETER satellite. Six examples of absolute flux measurements
    are given in units of electrons  cm$^{-2}$s$^{-1}$sr$^{-1}$keV$^{-1}$ during wisp events.
     The signal is adopted from data recorded in April 2007, when the NWC transmitter was operating,
      whereas the background data were obtained in July 2007, when the NWC transmitter was deactivated.
  For example, the first plot is the effect of the signal data in the NWC electron precipitation
   area on April 14th. Thus, the background data recorded in the same area on July 14th have been subtracted. }
  \label{3}
 \end{figure}

Examples of NWC wisp structures observed by the DEMETER satellite
are exhibited in Fig.\ref{3}, which displays various NWC electron precipitation belts
 obtained during NWC transmitter operation.
  Note that the background data when the NWC transmitter was deactivated
   are subtracted from these results.

\section{Verification of Quasi-linear Diffusion Equation in Wave-Particle Interaction Theory}  

\subsection{Quasi-linear Models}
Originating from the Fokker-Planck equation, the quasi-linear
diffusion equation has been developed as a practical and
convenient form\cite{Schulz1974,Tao2008,Subbotin2009,Shprits2009b,Su2010}. Similar to
the dipole field case, the diffusion equation used in this paper,
 which includes both pitch angle and momentum diffusion, is expressed as

\begin{eqnarray}
\frac{\partial f}{\partial t}&=&
\frac{1}{T(\alpha)}\frac{\partial}{\partial\alpha}[T(\alpha)(<D_{\alpha\alpha}>\frac{\partial
f}{\partial\alpha}+<D_{\alpha p}>\frac{\partial f}{\partial p})]
\\
&&+\frac{1}{p^2}\frac{\partial}{\partial
p}[p^2(<D_{p\alpha}>\frac{\partial
f}{\partial\alpha}+<D_{pp}>\frac{\partial f}{\partial p})]-\frac{f}{\tau},
\end{eqnarray}
where
\begin{equation}
T(\alpha)=S(\alpha)\sin2\alpha.
\end{equation}
Here, $S(\alpha)$ is a function corresponding to the bounce
frequency obtained from the bounce-averaged performance, which is
estimated in a dipole field\cite{Lenchek1961} such that
\begin{equation}
S(\alpha)=1.38-0.32(\sin\alpha+\sqrt{\sin\alpha}).
\end{equation}
$\tau$ is the electron lifetime, which is generally set to be a quarter of the bouncing time in the
loss cone and $f/\tau$ is infinite outside the loss cone.
The $<D_{\alpha\alpha}>$, $<D_{pp}>$, and $<D_{p\alpha}>$ terms correspond to the bounce-averaged
 diffusion coefficients for the pitch angle diffusion, momentum diffusion, and their mixed term, respectively,
  having the following detailed formulae:

\begin{equation}
<D_{\alpha\alpha}>=\frac{1}{S(\alpha_{eq})}\int^{\lambda_m}_{0}d\lambda
D_{\alpha\alpha}\frac{\cos\alpha}{\cos^2\alpha_{eq}}\cos^7\lambda,
\end{equation}

\begin{equation}
<D_{\alpha
p}>=\frac{1}{S(\alpha_{eq})}\int^{\lambda_m}_{0}d\lambda D_{\alpha
p}\frac{(1+3\sin^2\lambda)^{\frac{1}{4}}}{\cos\alpha_{eq}}\cos^4\lambda,
\end{equation}

\begin{equation}
<D_{pp}>=\frac{1}{S(\alpha_{eq})}\int^{\lambda_m}_{0}d\lambda
D_{pp}\frac{(1+3\sin^2\lambda)^{\frac{1}{2}}}{\cos\alpha}\cos\lambda,
\end{equation}
where $\lambda_m$ denotes the latitude of the bounced-particle mirror point.
 Therefore, the phase space density is regarded as a function of equatorial pitch angle $\alpha_{eq}$, the kinetic energy $E$,
   and the L-shell with the expression $f(\alpha_{eq},E,L)$.

 The wave-frequency spectrum density is assumed to obey a Gaussian distribution. Assuming the field-aligned electromagnetic wave, the closed analytical form of the local pitch-angle diffusion coefficient of the wave-particle interaction is derived as follows (see Ref.\cite{Summers2005}, Eqs. (33)--(35)):
\begin{eqnarray}
D_{\alpha\alpha} & =&
\frac{\pi}{2}\frac{1}{\nu}\frac{\Omega^2_{\sigma}}{|\Omega_e|}\frac{1}{(E+1)^2}\sum_s\sum_j\frac{R(1-\frac{x\cos\alpha}{y\beta})^2|F(x,y)|}{\delta
x|\beta \cos\alpha-F(x,y)|} \nonumber\\
&& \cdot  e^{-(\frac{x-x_m}{\delta x})^2},
\end{eqnarray}
\begin{eqnarray}
\frac{D_{\alpha p}}{p} & =&
\frac{\pi}{2}\frac{1}{\nu}\frac{\Omega^2_{\sigma}}{|\Omega_e|}\frac{\sin\alpha}{\beta(E+1)^2}\sum_s\sum_j\frac{R(\frac{x}{y})(1-\frac{x\cos\alpha}{y\beta})|F(x,y)|}{\delta
x|\beta \cos\alpha-F(x,y)|} \nonumber\\
&& \cdot  e^{-(\frac{x-x_m}{\delta x})^2},
\end{eqnarray}
\begin{eqnarray}
\frac{D_{pp}}{p^2} & =&
\frac{\pi}{2}\frac{1}{\nu}\frac{\Omega^2_{\sigma}}{|\Omega_e|}\frac{\sin^2\alpha}{\beta^2(E+1)^2}\sum_s\sum_j\frac{R(\frac{x}{y})^2|F(x,y)|}{\delta
x|\beta \cos\alpha-F(x,y)|} \nonumber\\
&& \cdot  e^{-(\frac{x-x_m}{\delta x})^2},
\end{eqnarray}
where $E$ is the dimensionless particle kinetic energy given by
$E=E_k/(m_{\sigma}c^2)=\gamma-1$,
$\beta=\nu/c=[E(E+2)]^{1/2}/(E+1)$, $R=|\delta B_s|^2/B^2_{0}$ is the ratio of the energy density of the turbulent magnetic field to that of the background field, and $B_0$ is the Earth's magnetic field. Further, $x_m=\omega_m/|\Omega_e|$, $\delta x=\delta
\omega/|\Omega_e|$, and $s=1$ for the R-mode wave, while $s=-1$ for the L-mode wave. $j=1,2,\cdots,N $ is the root number satisfying the resonance
condition. We have
\begin{equation}
\omega_j-\nu \cos\alpha
k_j=-s\frac{q}{|q|}\frac{|\Omega_{\sigma}|}{\gamma}.
\end{equation}
$F(x,y)=dx/dy$ ($x=\omega/|\Omega_e|, y=ck_i/|\Omega_e|$) is
determined from the dispersion equation of the electrons or protons as
follows:
\begin{equation}
(\frac{ck}{\omega})^2=1-\frac{(1+\epsilon)/\alpha^{\ast}}{(\omega/|\Omega_e|-s)(\omega/|\Omega_e|+s\epsilon)},
\end{equation}
where
\begin{equation}
\alpha^{\ast}=\Omega^2_e/\omega^2_{pe},
\end{equation}
is an important cold-plasma parameter; $\epsilon$ is the rest mass
ratio of an electron and proton; $|\Omega_e|=e|B|/(m_ec)$ denotes
the electron gyrofrequency, and $\omega_{pe}=(4\pi
N e^2/m_e)^{1/2}$ is the plasma frequency, where $N$ denotes
the particle number density in the ionosphere and is determined by formula $N=N_0 \times (2/L)^4 cm^{-3}$,
        according to Refs.\cite{Angerami1964,Inan1984}.
\subsection{Numerical Method, Initial and Boundary Conditions}

We solve the bounce-averaged Fokker-Planck equation numerically on ($\alpha_{eq}$, $p$) plane at different $L$.
The domain of our simulation corresponds to $[0, \frac{\pi}{2}]$ in $\alpha_{eq}$, $[0~\text{MeV}, 1.5~\text{MeV}]$ in $E_{k}$, and $[1, 3]$ in $L$.
The grid points are chosen to be $200 \times 200 \times 100$ for equaiorial pitch angle, momentum, and $L$-shell, respectively.
And these grid points are distributed uniformly in our simulation domain.
Our time step in this simulation is 100~s.

Since mixed partial derivatives appear in our diffusion equation, we choose fully implicit finite difference
 method to keep our simulation stable (similar to \cite{Subbotin2009}).
Schematically, our equation takes the form
\begin{equation}
    \frac{\partial f}{\partial t} = \boldsymbol{F} f
\end{equation}
where $\boldsymbol{F}$ represents the differential operator on the right-hand-side.
The implicit finite difference method can be written as
\begin{equation}
    \frac{f^{n + 1} - f^{n}}{\Delta t} = \boldsymbol{F} f^{n + 1}
\end{equation}
where $n$ is the time index.
By solving the linear equation
\begin{equation}
    (\boldsymbol{I} - \Delta t \boldsymbol{F}) f^{n + 1} = f^{n}
\end{equation}
at each step, we can obtain numerical solution of $f$ at arbitrary time.
To reduce solving time at each step, we apply LU decomposition to the discretized equation before simulation starts.

In order to resolve the wave particle interaction equations with mixed derivative term, many numerical simulation methods are introduced (\cite{tHout2010,Su2010}). Here our simulation time is only several hours, with the time step 100 s, so a conventional fully implicit finite difference method is enough in this work.

In the numerical simulation of the quasi-linear diffusion to the NWC electron belts, the initial diffusion function $f_{0}$ satisfies the empirical formula
\begin{equation}
f_0=\sin\alpha_{eq}\cdot e^{-\frac{E-0.1}{0.2}}
\end{equation}

In our simulation, we apply Dirichlet boundary condition at $E_{k} = 0~\text{MeV}$,
and Neumann boundary condition on other boundaries.
For $\alpha_{eq}$
\begin{equation}
    \left. \frac{\partial f}{\partial \alpha_{eq}} \right|_{\alpha_{eq} = 0} = 0 ,
    \qquad
    \left. \frac{\partial f}{\partial \alpha_{eq}} \right|_{\alpha_{eq} = \frac{\pi}{2}} = 0
\end{equation}
and for $E_{k}$
\begin{equation}
    \left. f \right|_{E_{k} = 0~\text{MeV}} = \left. f_{0} \right|_{E_{k} = 0~\text{MeV}} ,
    \qquad
    \left. \frac{\partial f}{\partial E_{k}} \right|_{E_{k} = 1.5~\text{MeV}} = 0
\end{equation}

\subsection{Simulation Results}
In order to simulate the NWC electron belts induced by whistler
waves in the ionosphere, the following parameters are chosen for use
 in the quasi-linear wave-particle coupling model: a Gaussian wave
  spectrum center frequency $\omega/(2\pi)=19.8$ kHz; a wave amplitude $\delta b= 200$ pT,
   which is consistent with the calculated result obtained via the NWC EM wave
    propagation described by full-wave model analysis \cite{Lehtinen2009,Zhao2015}, a semi-bandwidth of 500 Hz,
    which is approximately consistent with the observed values obtained from the electric-field
     instruments on board the DEMETER satellite, and an equatorial magnetic field
      $B=3.11\times 10^{-5}/L^3 $ T, which is determined by a dipole model.

We still consider the effect of the bounce loss cone and
drift loss cone in the
numerical simulation on the NWC electron precipitation belts.
 The bounce loss cone $LC_{bounce}$ depends on the L-shell
according to the expression
\begin{equation}
LC_{bounce}=\arcsin{\sqrt{\frac{(\frac{6470}{6370L})^3}{\sqrt{4-3(\frac{6470}{6370L})}}}}.
\end{equation}
In theory, drift loss cone $LC_{drift}$ varies with longitude; however, it is actually affected by many factors, such as the season or certain magnetic storms. In the local area above the NWC transmitter position, we will exhibit the effect of $LC_{drift}$ value at approximately $5^{\circ}$, $10^{\circ}$ and $15^{\circ}$, respectively. The electron lifetime $\tau$ is generally set to be a quarter of the bouncing time in the loss cone and infinite outside the loss cone. But in inner radiation belt, the bounce loss cone is larger and the observed electrons by satellite come from precipitated electrons in loss cone. $\tau$ with too small value will induce too weak effect of electron precipitation in certain time. So here we set $\tau$ value to be 60 s.

\begin{figure}
\centering
\includegraphics[width=38pc]{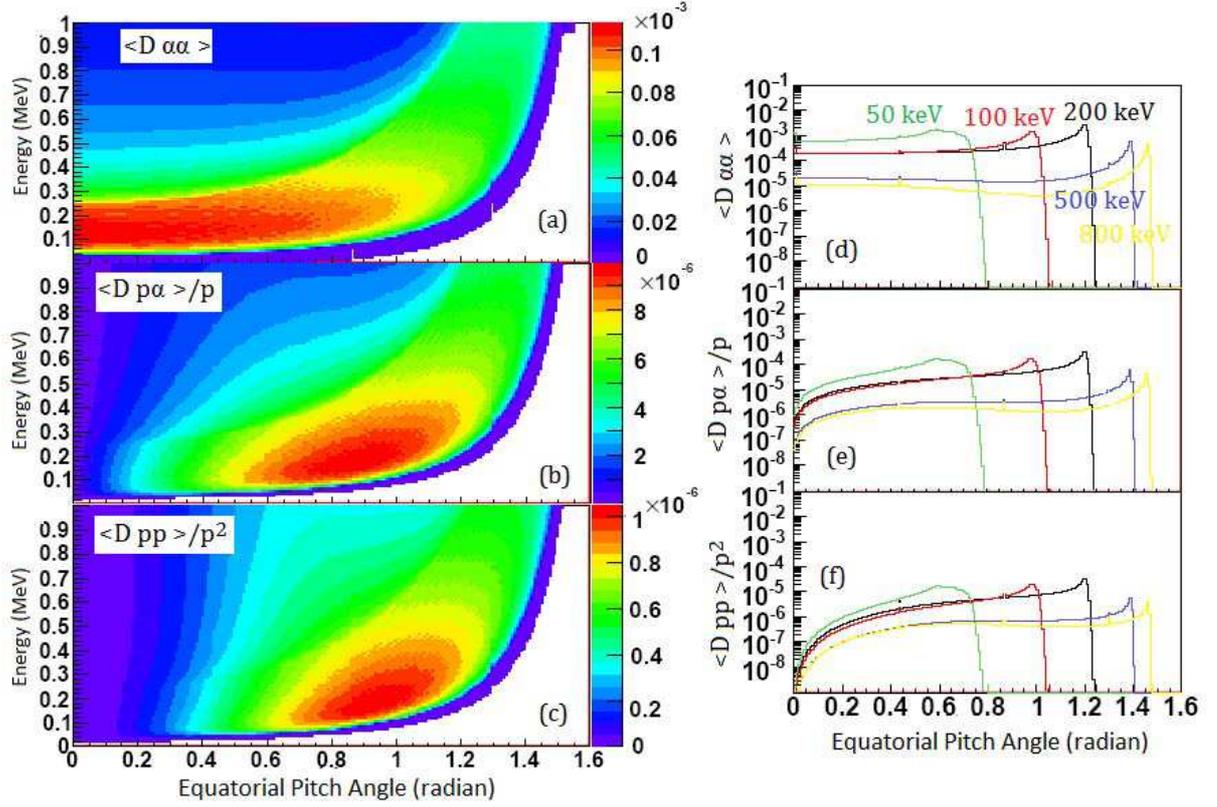}
  \caption{\small Diffusion coefficient distributions with parameters: wave frequency $\omega/(2\pi)=19.8$ kHz, wave amplitude $\delta b=200$ pT, semiband width 500 Hz, and $N_0=3000$ counts/s. L-shell  in (a) (b) (c): L = 1.4--2.0, L-shell in (d) (e) (f): L= 2. Five colorful lines denotes the diffusion coefficient for electrons with 50 keV, 100 keV, 200 keV, 500 keV and 800 keV. }
  \label{4}
 \end{figure}

  The bounce-averaged Diffusion coefficients are exhibited in Fig.\ref{4}. We can see that
  the pitch angle diffusion is dominant and larger by about two orders of magnitude than momentum and mixed diffusion terms. And pitch angle diffusion for electrons in L = 1.4 -- 2.0 plays a strongest role in energy range of 0.05 -- 0.25 MeV.

\begin{figure}
\centering
\includegraphics[width=38pc]{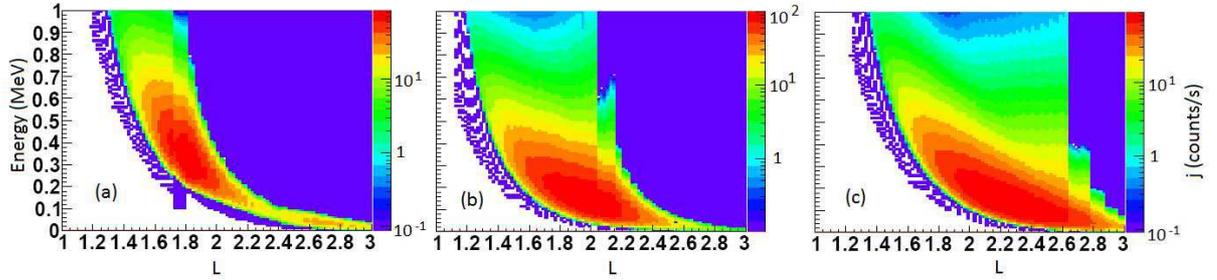}
  \caption{\small Simulated NWC electron precipitation structures
  after subtraction of initial flux distribution from final flux distribution.
   Parameters used in simulation: frequency $\omega/(2\pi)=19.8$ kHz; wave amplitude $\delta b=200$ pT; semiband width 500 Hz;
   and $N_0=1000$ counts/s. The equatorial pitch angles $\alpha_{eq}$
    used in the simulation ranged from 0.3 to 0.44, corresponding to the observation range by the DEMETER and NOAA satellites.
      The loss cones were set to:
   $LC_{drift} =$ (a) $5^{\circ}$; (b) $10^{\circ}$; (c) $15^{\circ}$, all of which were with a bounce loss cone. }
  \label{5}
 \end{figure}

Following the above numerical simulation method in section 3.2,
simulated results for 4-h evolution were obtained and are
presented in Fig.\ref{5}. These results indicate the relatively good
agreement with the satellite observation for the electron flux
distribution, which is shown as a function of energy and L-shell (see Fig.\ref{3}).
Identical parameters to those listed above were used in the
simulation. Eqs. (1) and (2) were solved in the radial distance
range from 1--3 earth radii ($R_E$) and for equatorial pitch angle
$\alpha_{eq}$ = 0 -- $\pi/2$. The energy range does not vary
with radial distance and was set to 0.05--1 MeV. As in the inner
radiation belt, the radial diffusion scattering plays a small role
in the total particle diffusion behavior. Note that all the plots
in this figure correspond to $\alpha_{eq} = 17$--$25^{\circ}$
(that is 0.3 to 0.44 radian), which is consistent with detection
range of the DEMETER satellite\cite{Li2012}. In the results shown
in Fig.\ref{5}, the particle distribution depends on the
$\sin\alpha_{eq}$; this is similar to the flat distribution inside
the loss cone adopted in Ref.\cite{Kennel1969}. We still
consider both the bounce and drift loss cone effects with $LC_{drift}$ set to be:
(a) $5^{\circ}$, (b) $10^{\circ}$, and (c) $15^{\circ}$.  The bounce loss cone is
determined by Eq. (14). From Figure 4, it can be seen that electron
precipitation also takes place for the ~50--100-keV range at higher L shell; this
result is in agreement with the observations of the NOAA
satellites.

\begin{figure}
\centering
\includegraphics[width=22pc]{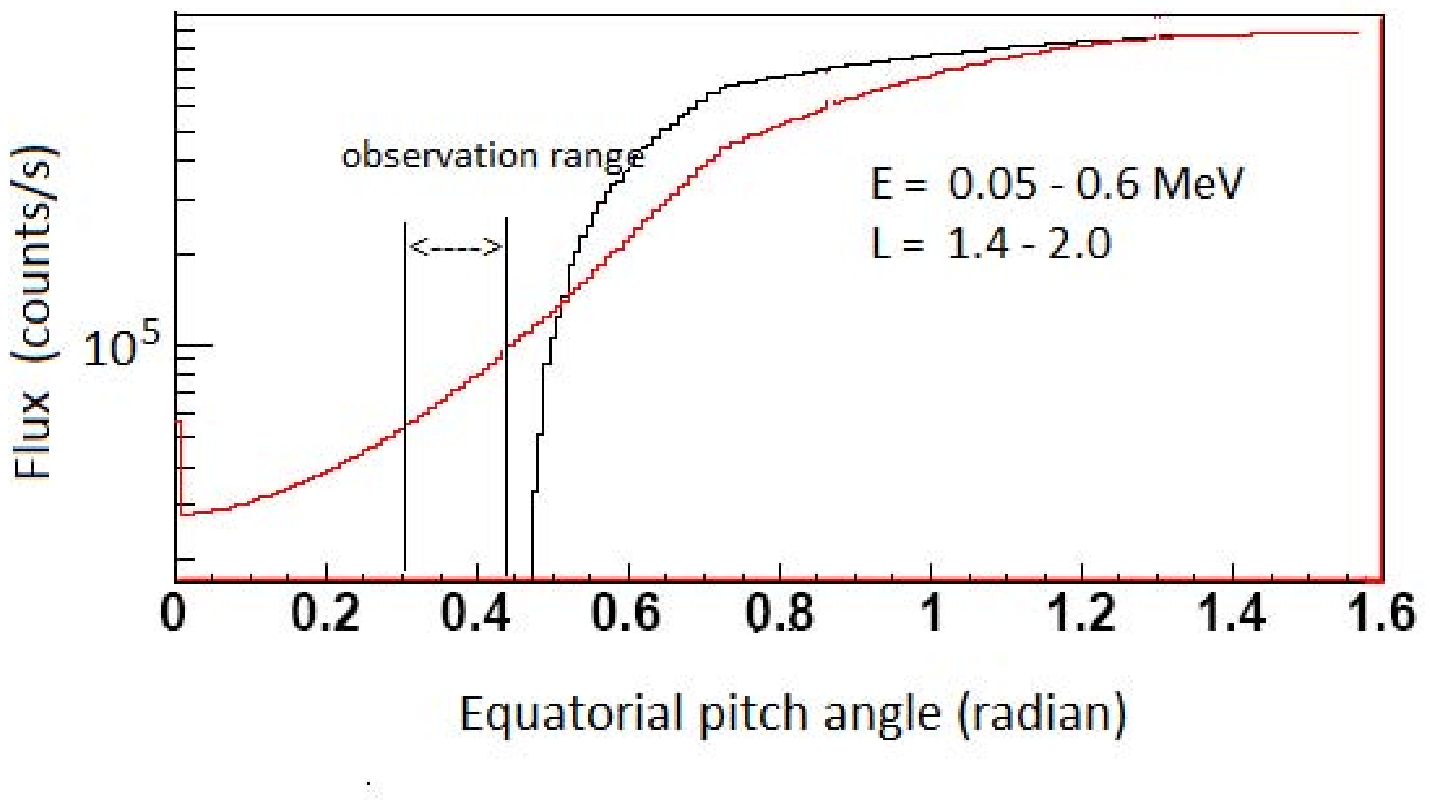}
  \caption{\small Flux distribution of initial and final states during wave particle interaction, corresponding to electrons with energy 0.05 -- 0.6 MeV and L shell value 1.4--2.0. Other parameter values: wave frequency $\omega/(2\pi)=19.8$ kHz, wave amplitude $\delta b = 200$ pT, semiband width 500 Hz, and $N_0=3000$ counts/s. The DEMETER and NOAA satellite observation ranges from 0.3 to 0.44 radian. }
  \label{6}
 \end{figure}

Fig.\ref{6} is the flux distribution of initial( black ) and final ( red ) states during wave particle interaction, corresponding to electrons with energy 0.05--0.6 MeV and L shell value 1.4--2.0. We can see that the satellite-observed electrons locate inside of loss cone when the electrons are scattered into loss cone mainly by pitch angle diffusion mechanism.

\begin{figure}
\centering
\includegraphics[width=40pc]{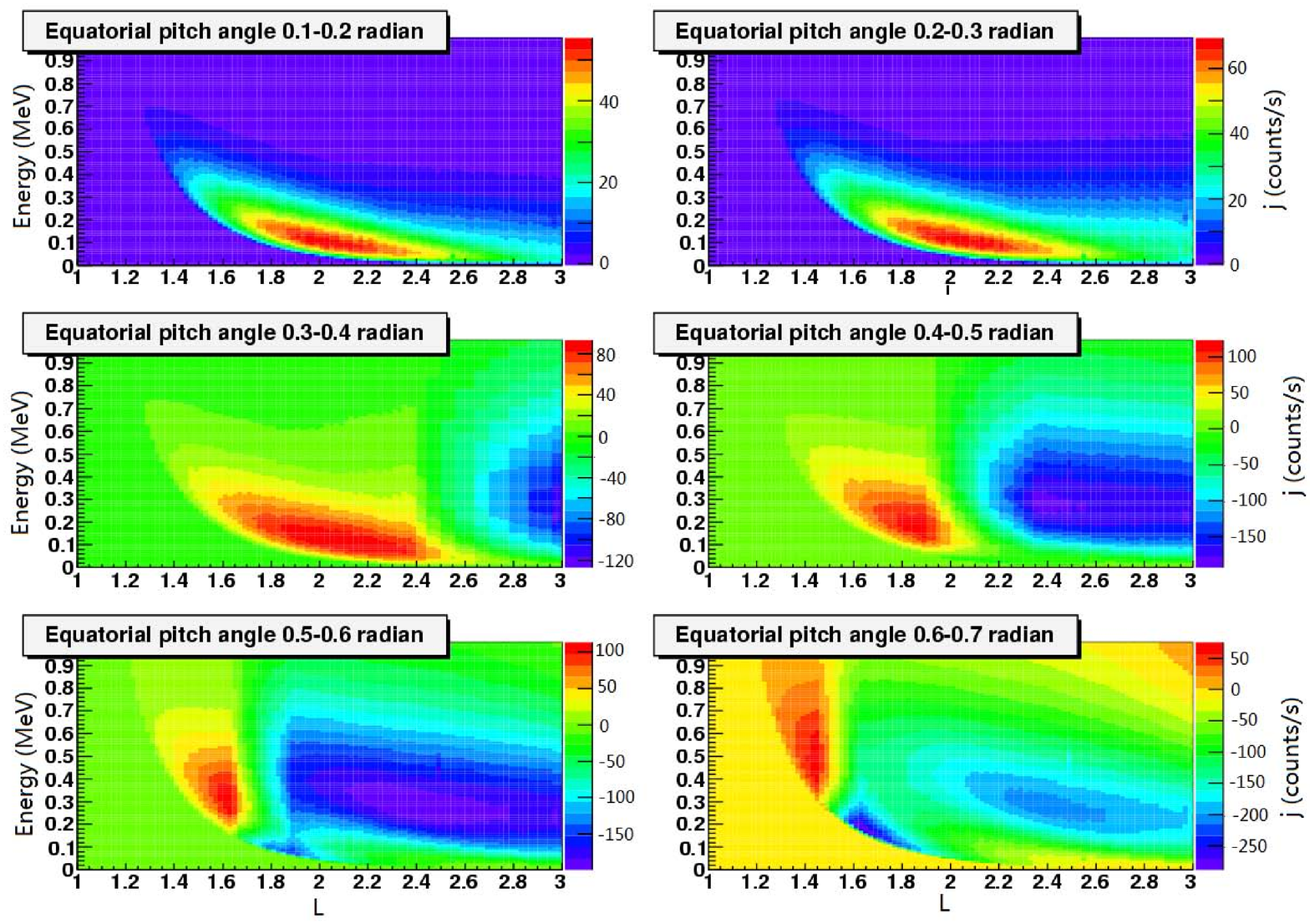}
  \caption{\small Simulated electron precipitation structures after subtracting initial flux distribution
   from final flux distribution. Simulation parameters: Electromagnetic wave
  frequency $\omega/(2\pi)=19.8$ kHz; wave amplitude $\delta b=200$ pT; semi-band width 500 Hz;
  $N_0=1000$ counts/s; and equatorial pitch angles $\alpha_{eq} =$ 0.1--0.6 radian. }
  \label{7}
 \end{figure}

Fig.\ref{7} is the simulated NWC electron precipitation for different equatorial pitch angle $\alpha_{eq}$ values of $0.1$--$0.2$, $0.2$--$0.3$, $0.3$--$0.4$, $0.4$--$0.5$, and $0.5$--$0.6$ radian. It is apparent that the electron precipitation induced by the wave-particle interaction differs in different $\alpha_{eq}$ ranges. The simulation result obtained for an $\alpha_{eq}$ of $0.3$--$0.4$ is located inside the IDP observation range on board the DEMETER satellite. Note that, at a higher $\alpha_{eq}$ range, it is easier for higher energy electrons to interact with VLF waves in the ionosphere. However, due to the pitch angle observation limitations of high-energy particle detectors and the lower density distribution of higher-energy electrons in space, it is possible that the electron precipitation induced by VLF waves is not observed by satellites to some extent.

\section{Asymmetric Distribution of the NWC Electric Field}
\begin{figure}
\centering
\includegraphics[width=40pc]{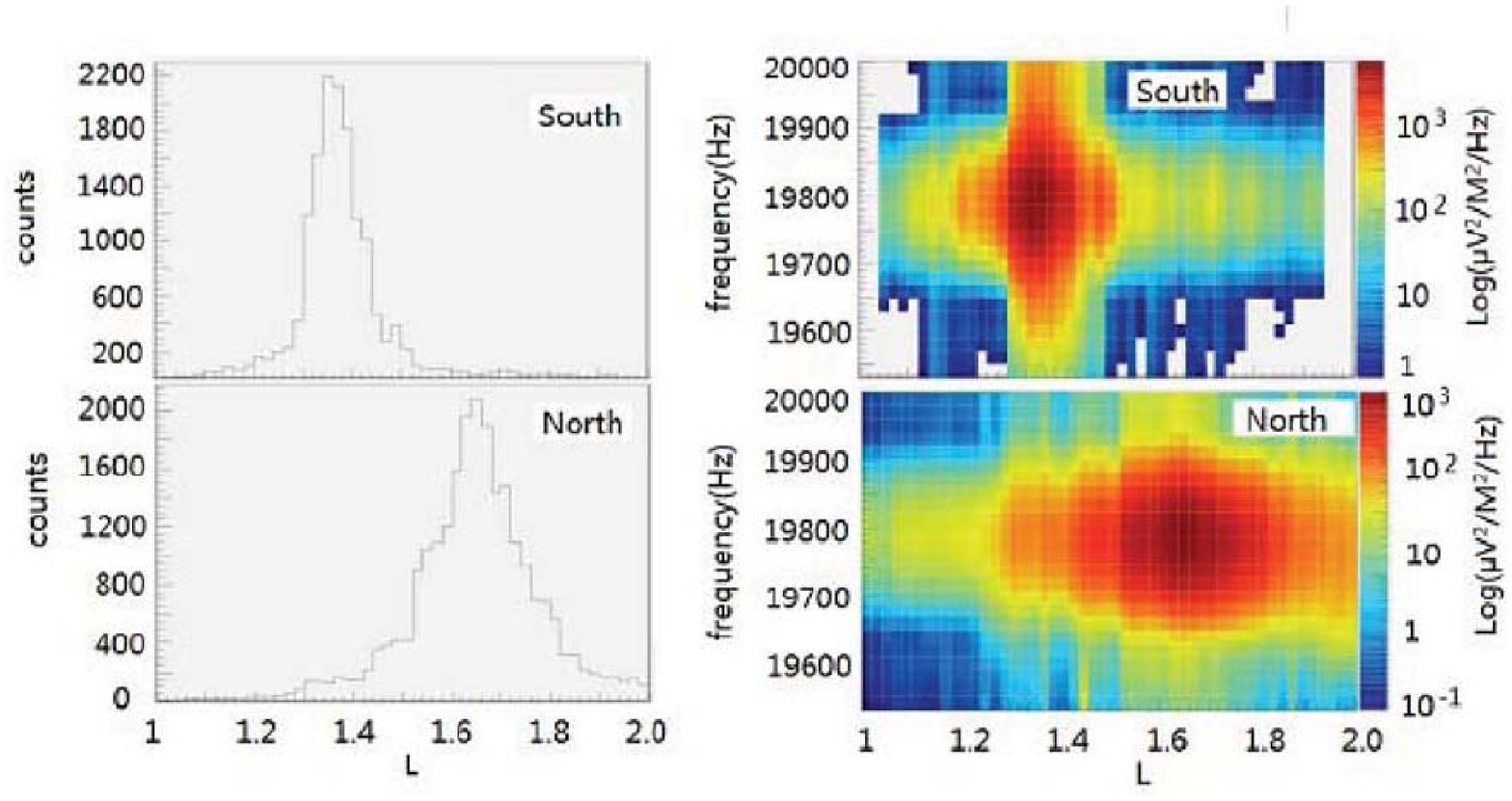}
  \caption{\small NWC electric-field power distributions measured
 by DEMETER satellite in January 2007. }
  \label{9}
 \end{figure}

The NWC electric field power distribution for the entirety of
January 2007, as detected by the DEMETER satellite,
is shown in Fig.\ref{9}. This figure shows that, at the altitude of
the DEMETER satellite orbit, the L-shell of the NWC electric-field
power covers $1.3$--$1.5$ and $1.5$--$1.8$ in the southern and northern
hemispheres, respectively. This phenomenon of asymmetrical
distribution of the NWC electric field in space is obvious.

\begin{figure}
\centering
\includegraphics[width=22pc]{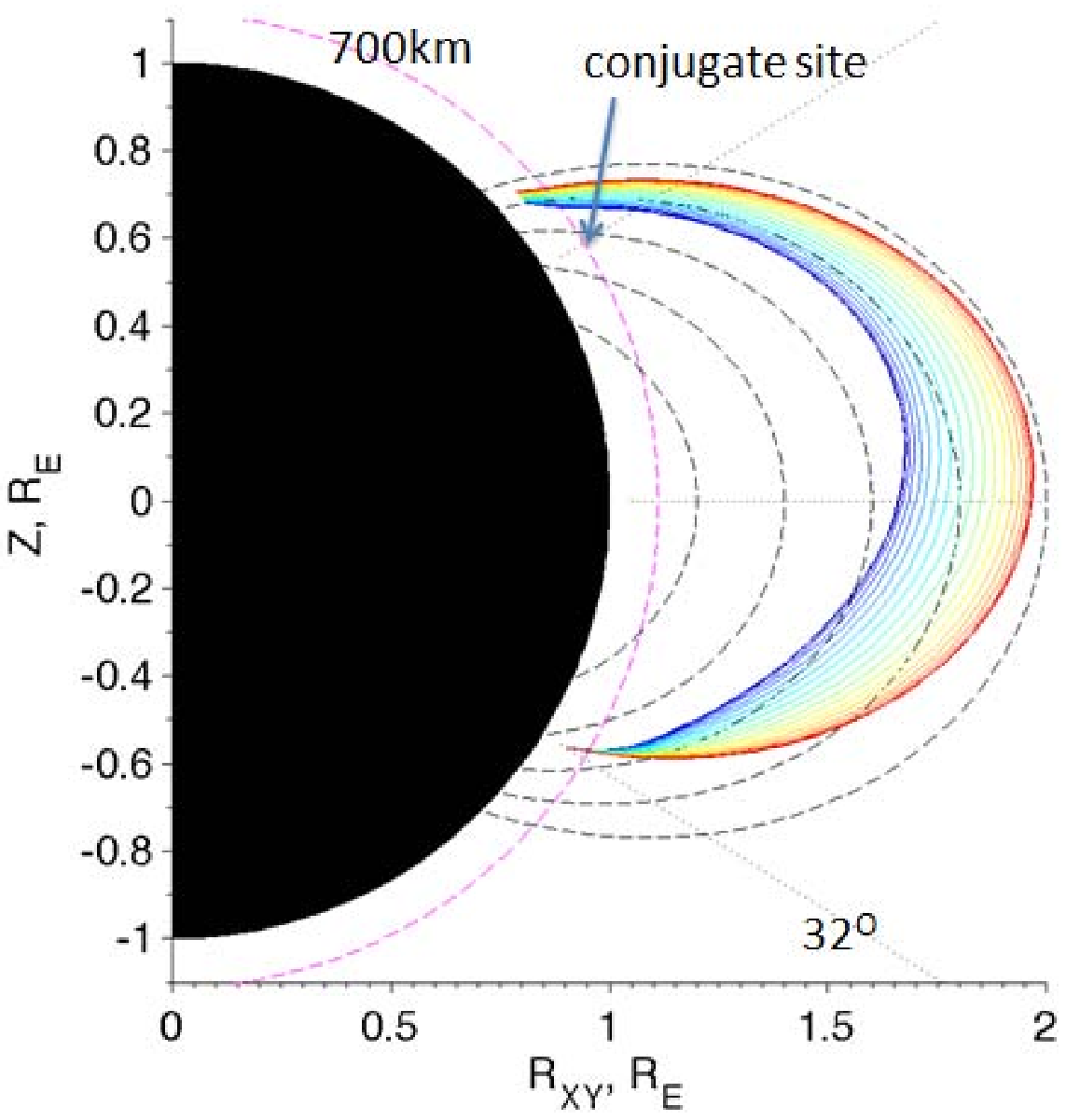}
  \caption{\small Simulated NWC transmission path by ray tracing model. Electromagnetic wave with 19.8 kHz is launched over a transmission cone of 5 deg with respect to vertical direction. }
  \label{10}
 \end{figure}

In order to explain this phenomenon, we use ray tracing model in a cold plasma in dipole magnetic field\cite{Horne1989,Chen2009, Chen2013,Xiao2007}, to simulate the NWC wave propagation path in magnetosphere.
Electromagnetic waves with 19.8 kHz are launched at L=1.35 close to NWC site on southern hemisphere, over a cone of 5 degree with respect to the vertical direction of earth's surface. The launch position is at geographic coordinate ($21.82^o$ S, $114.15^o$ E) and geomagnetic coordinate ($-31.96^o$, $186.4^o$).
Fig.\ref{10} shows the simulated ray propagation pathes. We can see that the L-shell values of NWC waves on north hemisphere at the conjugate latitude is higher. This asymmetry nature of NWC wave in southern and northern hemispheres is caused by the refraction effect of wave propagation in the inhomogeneous magnetosphere.

\section{Discussion and Summary} 

In this work, in addition to the previously reported electron precipitation with 100--600-keV energy
 detected by the DEMETER satellite, we have still reported enhancement of the 30--100-keV electron flux
  observed by the NOAA satellites at night, when the powerful NWC VLF transmitter is broadcasting.
   We then performed a validity check using a theoretical model, i.e., the quasi-linear diffusion equation,
    and demonstrated approximate agreement with the satellite observation results.

\begin{figure}
\centering
\includegraphics[width=40pc]{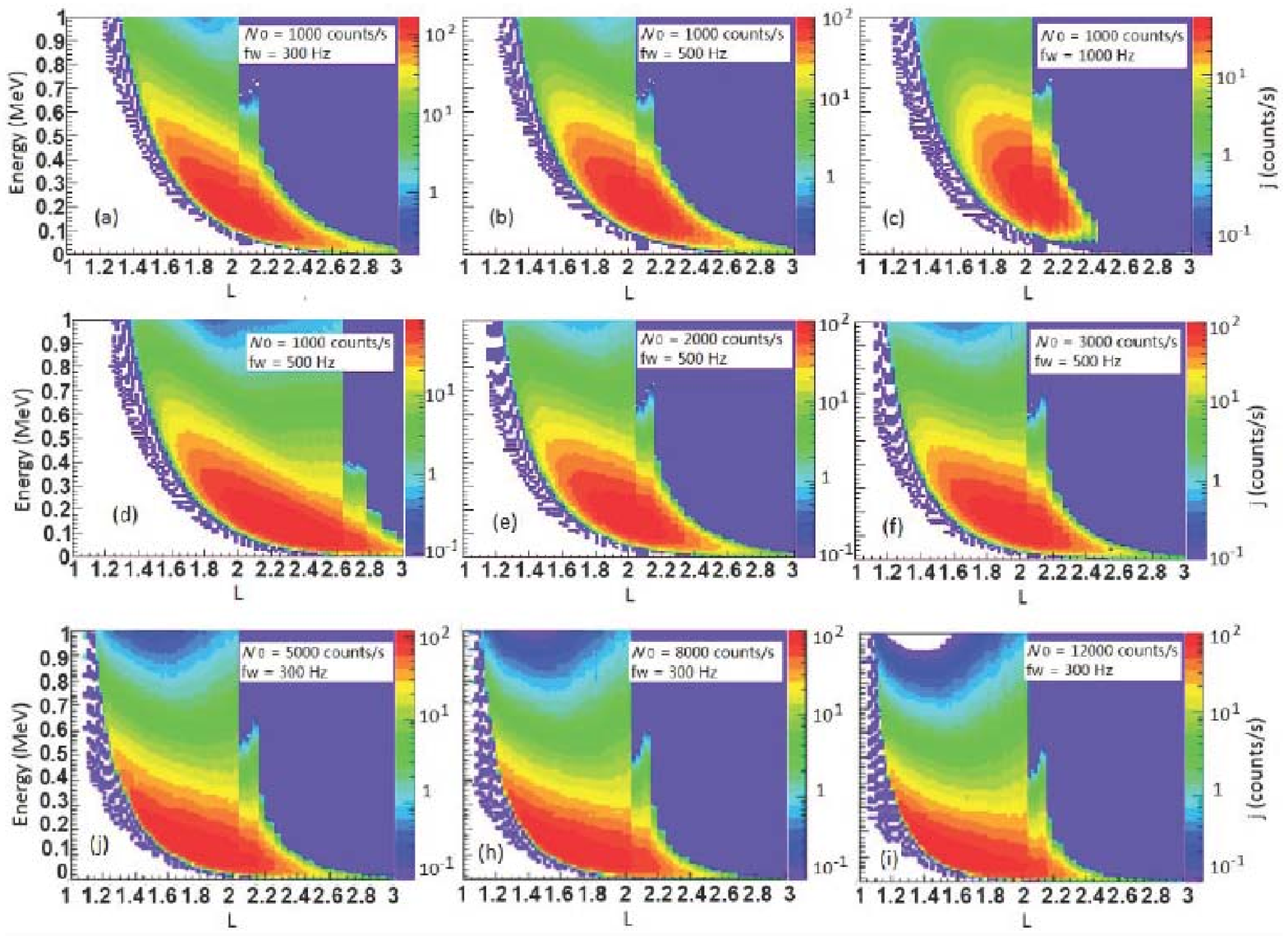}
  \caption{\small Simulated electron precipitation structures after subtracting initial flux distribution
  from final flux distribution. Simulation parameters: Electromagnetic wave
  frequency $\omega/(2\pi)=19.8$ kHz; wave amplitude $\delta b=200$ pT.
    The equatorial pitch angles $\alpha_{eq}$ range from 0.3 to 0.44, which is consistent with the DEMETER observation. }
  \label{8}
 \end{figure}

Many factors influence the simulation results obtained using the quasi-linear diffusion equation,
 including the adopted wave semi-bandwidth, electron density. In Fig.\ref{8}, we show the simulated electron precipitation
  structure obtained for wave semi-bandwidths of 300, 500 and 1000 Hz and different electron densities. It is apparent
   that the simulated wisp structure is wider for a greater wave width. The simulated wisp structure moves toward to the lower L value with the increasing electron density. Thus,
   the parameter uncertainty, including loss cone, wave width and electron density, may consist of the factors which cause the difference of simulated result with the real wisp shapes observed via satellite.

Thus, these results indicate that the quasi-linear wave particle interaction model can describe
 an NWC electron precipitation wisp structure approximately, and can also explain the coupling
  characteristics of wave-particle interaction in a general sense. In particular,
   below the 0.1-MeV electron energy, the theoretical simulation results demonstrate
    agreement with the NOAA satellite observations shown in Fig.\ref{2}.

Electron precipitation or accumulation in the ionosphere may also be
caused by both energy diffusion and radial diffusion in the
L-shell\cite{Falthammar1965,Su2015}. However, in the inner
 radiation belt, the effect of radial diffusion on the L-shell can be omitted, especially
  at the satellite altitudes of hundreds of kilometers. The radial diffusion plays a large role in the wave-particle interaction only in the outer radiation belt \cite{Su2011}. In addition, when the theoretical simulation time is sufficiently long, e.g., having a duration of
   approximately 1 yr\cite{Selesnick2013}, the radial diffusion in the L-shell evolution
 must be considered in the wave-particle interaction analysis in the inner radiation belt.
 Thus, significant attention is not paid to the effect of
    radial diffusion on the NWC electron belts in the analysis of the wave-particle interaction
    simulation conducted in this study.

However, we must note that the quasi-linear theory is thought to
describe interactions between charged particles and
small-amplitude broadband waves\cite{Kennel1966}. When the wave
amplitude increases or the band becomes narrower, nonlinear
effects such as phase trapping and bunching are more likely to
become dominant, calling the use of quasi-linear theory into
question\cite{Bortnik2008}. Inan {\it et al.} have studied the
nonlinear pitch angle scattering of energetic electrons by
coherent VLF waves in the magnetosphere and compared their
observations with linear theory\cite{Inan1978}. They have also
defined a quantity $\rho$, the ratio of the maximum absolute
values of the wave and inhomogeneity terms, so as to differentiate
between the linear and nonlinear interactions. Furthermore, Tao
{\it et al.} have reported that the nonlinear interactions may be
important near the equatorial plane, even for a moderate wave
amplitude, and have shown that phase trapping is likely to occur
when the wave-induced motion dominates the adiabatic
motion\cite{Tao2010}. Those researchers subsequently used a
test-particle code to confirm that the effect of the amplitude
modulations should be considered in the quantitative treatment of
nonlinear interactions between electrons and chorus
waves\cite{Tao2012,Tao2013}. There are other related works about
dynamics of high-energy electrons interacting with whistler mode
chorus emissions studied by test particle
simulations\cite{Hikishima2009,Omura2006}. Su {\it et al.} pointed
out that the nonlinear physical processes, including boundary
reflection effect, phase bunching and phase trapping, start to
occur as the amplitude increases by studying the interaction
between electrons and parallel-propogating monochromatic EMIC
waves\cite{Su2012}. And then he further presented that the
nonlinear processes depend on the electron initial latitude and
decrease or increase the loss rate predicted by the quasi-linear
theory\cite{Su2013}.
In addition, the azimuthal advection may play an important role in the radiation belt electron dynamics\cite{Su2010}, in the next work, we'll study this effect and attempt to incorporate it in theory model for research of ground-based induced electron precipitation belts.

Therefore, in our next work, we will attempt to adopt a non-linear
theory model to further study this wave-particle interaction
event, focusing on the quantitative calculation.

\begin{figure}
\centering
\includegraphics[width=22pc]{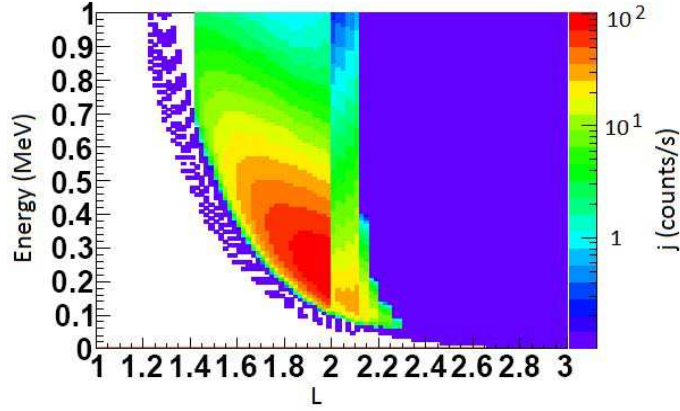}
  \caption{\small Simulated NWC electron precipitation structure
  following subtraction of initial flux distribution from final flux distribution.
   The NWC electric-field power distributions in the L-shell in the south
    and north hemispheres are considered in the simulation process.
     Simulation parameters: Electromagnetic wave frequency $\omega/(2\pi)=19.8$ kHz;
      wave amplitude $\delta b=200$ pT; semi-bandwidth 500 Hz; $N_0=1000$ counts/s.
       The equatorial pitch angles $\alpha$ used in the simulations range from 0.3 to 0.4,
        which is consistent with the value of the NWC electron precipitation wisp structures
         observed by the DEMETER and NOAA satellites. The bounce loss cone and $LC_{drift} = 5^{\circ}$
          are set in the simulation. }
  \label{11}
 \end{figure}

Also considering the real NWC electric-field power distributions in the L-shell for the south and north hemispheres observed by the DEMETER satellite, shown in Fig.\ref{9}, the wave-particle interaction is simulated numerically
    using the quasi-linear diffusion equations described above. The simulation results are shown in Fig.\ref{11}.

From the simulation results shown in Fig.\ref{8}, the distribution trend of the NWC electron
 precipitation in the L-shell and the energy dimensions remain consistent with the observed
  wisp structure, although the simulated result is less accurate. This deviation
   from the observation results may be caused by a number of factors,
   including the influence of the wave bandwidth and cold plasma density.
 Additionally, the wave propagation effect in the ionosphere must be further studied in the next research stage.
  A comprehensive theoretical model that incorporates the wave propagation effect (such as the ray tracing model)
   and wave-particle interaction will facilitate a more complete and accurate approach to describing
    the electron precipitation in the radiation belt induced by the ground-based NWC transmitter.

In addition, we believe that analysis of the effects of the NWC
VLF wave propagation and electron precipitation, along with
development of a comprehensive theoretical model, will facilitate
the study of ground-based waves propagating into the ionosphere
and interacting with high-energy particles in radiation belts.
Based on wave particle interaction theory, investigation of the EM disturbance in the ionosphere induced by
seismic electromagnetic signals will be particularly useful, supposing that it is possible for seismic electromagnetic
 signals to propagate into ionosphere. As regarding to the influence of the EM waves emitted
by seismic activities in wider space, a large number of works have
also been published on this topic, including studies based on
satellite observation\cite{Zhang2012} and theoretical
exploration research\cite{Zhang2014a,Zhang2014b}.

\vspace*{2mm}

\section{Acknowledgement}
The authors are grateful to Tao Xin, Zong Qiugang, Wang Yongfu for useful discussions.




\end{document}